# Simultaneous excitation of phonons at the center and boundaries of Brillouin zones with high energy electron beams

Kartik Venkatraman[1], Peter A. Crozier[*1]

[1] School for the Engineering of Matter, Transport and Energy, Arizona State University, Tempe, AZ, USA

*corresponding author: crozier@asu.edu

Abstract

High energy electron beams can now be routinely focused to 1-2 Å and offer the ability to obtain vibrational information from materials using monochromated electron energy-loss spectroscopy (EELS) in a scanning transmission electron microscope (STEM). Here it is shown that long and short wavelength phonons can be probed simultaneously with on-axis vibrational STEM EELS. The advantage of using focused electron beams is that the short wavelength phonons are probed via impact scattering while the long wavelength modes are probed via dipole scattering. The localized character of the short wavelength modes is demonstrated by scanning the electron beam across the edge of a hexagonal boron nitride nanoparticle. It is found that the condition required for high spatial resolution STEM imaging also correlates with the condition to enhance the short wavelength phonon contribution to the vibrational energy-loss spectrum. Probing short wavelength phonons at high spatial resolution with on-axis vibrational STEM EELS will help develop a fundamental connection between vibrational excitations and bonding arrangements at atomic scale heterogeneities in materials.

1. Introduction

Phonons are quantized crystal lattice vibrations which control all thermally activated processes including diffusion, heat transport, electrical properties, phase transformations, and surface chemistry. Their wavevectors can range from zero at the center up to the Brillouin zone boundary, and the relationship between wavevector and energy is typically expressed through phonon dispersion curves [1]. Nanoparticles and nanostructure associated with polar materials may cause changes in the long wavelength phonon character close to the center of the Brillouin zone via the creation of phonon polaritons [2]. Crystal defects including vacancies, interstitials, dislocations, and grain boundaries locally alter the short wavelength modes close to the Brillouin zone boundary (BZB), and change materials properties significantly. Widely available optical techniques such as Raman and infrared (IR) spectroscopy are highly effective at characterizing long wavelength phonons close to the center of the Brillouin zone. Measuring dispersion curves and short wavelength phonons close to the BZB is more challenging and typically relies on less common techniques such as inelastic neutron, x-ray, or helium scattering, or low-energy electron scattering [3–8].

These scattering techniques have been successful at characterizing bulk systems and macroscopic surfaces but are not suitable for high spatial resolution investigations on nanoparticles or at defects. Near field optical techniques have nanometer resolution, but as mentioned earlier, are limited to long-wavelength phonon modes at the center of the Brillouin zone [9,10]. High energy electron beams are potentially a useful characterization tool to quickly probe the short wavelength phonon modes at the BZB with high spatial resolution.

High energy electron beams can now be routinely focused to sub nanometer dimensions in the STEM making atomic resolution core-loss EELS routinely available [11]. Recent developments in electron monochromation have dramatically improved the energy resolution of EELS making vibrational spectroscopy now available in the STEM [12,13]. The strong dipole interactions associated with electrons provide a sensitive probe for long wavelength phonons at the center of the Brillouin zone yielding information that is similar to IR spectroscopy. Electrons are able to carry and transfer significant momentum to the system and thereby sense the short wavelength phonons at the BZB. Recently, this approach has been employed to perform momentum-resolved EELS to determine the phonon dispersion curves in a graphene monolayer and a boron nitride flake using electron beams sized 2 – 40 nm [14,15].

It would be very useful to be able to rapidly sample long and short wavelength phonons simultaneously. Here we show, using suitable on-axis detection geometry, that it is possible to acquire vibrational EELS spectra which show features corresponding to phonons at the center *and* boundaries of the Brillouin zone. The approach involves using a large incident convergence angle to provide a 1-2 Å electron probe and a large range of incident momenta to illuminate a hexagonal boron nitride (h-BN) nanoparticle tilted into a zone axis orientation. Excitation of long wavelength modes is associated with small scattering angles, while that of short wavelength modes is associated with large scattering angles (comparable to the Bragg angle). By recording spectra at small scattering angles, it is easy to identify long wavelength modes which can then be subtracted from the spectra recorded at large scattering angles to leave only the short wavelength modes from the BZB. These spectral features are the result of impact scattering of the electron beam and give a more localized response than the features associated with dipole scattering which are the long wavelength modes from the center of the Brillouin zone. The high degree of localization offers the potential to sample short wavelength modes with nanometer or atomic resolution.

2. Experimental

2.1. Specimen preparation

High purity h-BN powder (99.8% pure) with average lateral dimensions of 5 μm and thicknesses of 300 nm was purchased from US Research Nanomaterials Inc. The powder was ultrasonicated in electronic grade (99.999% pure) isopropyl alcohol (IPA), purchased from Sigma Aldrich, for 1 hour. Ultrasonication helps exfoliate the layered h-BN powder into thinner flakes. The ultrasonicated solution was drop-casted onto a 3 mm, lacey-carbon-film coated, 200 mesh Cu TEM grid purchased from Pacific Grid-Tech and the grid was dried under a heat lamp for 1 hour. The specimen was baked at 140°C for 12 hours prior to introduction into the microscope to burn off all volatile hydrocarbons and prevent contamination.

2.2 Monochromated STEM EELS measurements

STEM EELS analysis on the specimen was performed using the monochromated, aberration-corrected Nion UltraSTEM 100, operating at 60 kV accelerating voltage. It was routinely possible to make 1Å sized probes with high convergence angles due to aberration correction up to the fifth order [16], and monochromation enabled a routine energy-resolution of 10 meV [17]. Each h-BN flake was oriented into the [0001] zone axis before EELS acquisition. The probe convergence semi-angle ($\alpha$) was either 10 or 33 mrad, and spectrometer collection semi-angle ($\beta$) was either 10 or 40 mrad. Figure 1a shows a schematic diagram of the STEM EELS geometry employed. Figure 1b overlays circles representing the different convergence and collection angles on the Brillouin Zones of [0001] h-BN.

Energy-loss information was acquired either as 'point-and-shoot' spectra from different positions on the flake or as linescans across the flake edge (an energy-loss spectrum was recorded at every pixel along the linescan with the pixel size being 2Å). An energy-dispersion of 0.66 meV per channel was used to record the spectra. Parametric two-window background subtraction was performed to isolate spectral features from the background of the zero-loss peak (ZLP) tail and of energy-losses preceding the feature of interest using custom MATLAB codes [18]. The spectra were calibrated by setting the center of the saturated ZLP to 0 meV, and the error in measuring energy-loss was the channel width i.e. ± 0.66 meV. The background subtracted signal was expressed as a linear combination of Gaussians to quantify the contributions from different peaks in the vibrational spectrum. The flake thickness was estimated by performing low-loss EELS measurements with 10 mrad probe convergence, 57 mrad spectrometer collection and an energy-dispersion of 50 meV per channel (Figure S1). The inelastic mean free path, $\lambda$, for valence-shell inelastic scattering from h-BN was computed to be 75 nm at 60 kV [19]. Measurements were performed on flakes of different thickness and the results in the figures come from a flake approximately 50 nm in thickness (t/λ value ~ 0.68).

3. Results and discussion

3.1. Detecting Long Wavelength Modes from Dipole Scattering

A typical background subtracted vibrational energy-loss spectrum from a h-BN nanoflake recorded with α = 10 and β = 10 mrad, is shown in Figure 2a. The inset shows an annular dark field (ADF) image of the flake and the probe position relative to the sample edge. The spectrum shows a strong peak at 194 meV with an intense shoulder at 175 meV, and a weak peak at 100 meV. There is weak intensity on the lower energy-loss tail of the 175 meV shoulder. Phonon dispersion curves from bulk h-BN obtained using inelastic X-ray scattering spectroscopy have been reproduced in Figure 2b to aid in the interpretation of spectral features [6]. In the current electron-optical geometry with convergence and collection angle significantly smaller than Bragg angles, most of the spectral features can be identified by looking at flat parts in the dispersion curves close to the Γ point. Thus, the 194 meV peak and its 175 meV shoulder are associated with the bulk h-BN longitudinal optical (LO) and transverse optical (TO) phonon modes, respectively. The weak peak at 100 meV seems to be the out-of-plane optical (ZO) phonon mode in h-BN. Due to the typically small scattering angles involved in their excitation, these modes are associated mostly with dipole scattering. There is also very weak intensity on the lower energy-loss tail of the 175 meV shoulder, which appears consistent with BZB modes at the M and K point. Gaussians were fit to the spectrum (Figure 2a) based on phonon mode positions at the Γ, M and K points in the dispersion curves and it is observed that the sum of all Gaussians fits the experimental spectrum well.

In an electron-optical geometry where dipole vibrational modes close to the Γ point are strongly excited, features in the energy-loss spectrum can be described by approximating the collective response of h-BN to the fast electron's electric field with its optical dielectric function [20,2]. h-BN is an anisotropic layered material, with mixed ionic-covalent bonding in the B – N planes and van der Waal's forces between the B – N planes [21]. Thus, it has different dielectric functions along the (0001) plane (in-plane) and perpendicular to it (out-of-plane). When the specimen is oriented along the [0001] zone-axis, there is only weak contribution from out-of-plane modes to the spectrum due to the small probe convergence. The in-plane dielectric function is shown in Figure 2c [21]. The real part of the in-plane dielectric function is negative between 170 and 196 meV; this region is called the Reststrahlen band. These are also the TO and LO phonon mode positions derived from Kramers-Kronig analysis of infrared reflection spectra, showing good agreement with the Γ point values in the dispersion curves. The real part of the out-of-plane dielectric function is positive in this region. The opposite signs of the real part of in-plane and out-of-plane dielectric functions leads to an indefinite dispersion for electromagnetic waves propagating through h-BN [22]. [0001] h-BN nanoflakes can thereby support volume and edge-surface hyperbolic phonon polaritons (HPhPs) due to anisotropy, and a previous vibrational EELS study from h-BN suggests that the peaks at 175 and 194 meV correspond to the volume HPhP and the bulk LO phonon modes respectively [2].

The Gaussian peak fitting method provides a simple and convenient method to extract the dipole and non-dipole components of the spectrum. Gaussians associated with dipole modes (100, 175 and 194 meV peaks) were added to give the dipole spectrum in Figure 2d. The non-dipole contribution (labelled BZB) is ~30 times weaker than the dipole contribution.

3.2. Detecting Short Wavelength Modes from Impact Scattering

*Increasing the Collection Angle*

Increasing the amount of Bragg scattered electrons entering the spectrometer will enhance the non-dipole contribution and may result in greater spectral contributions from short wavelength modes at the BZB. To test this hypothesis, β was increased to 40 mrad which is significantly larger than the smallest Bragg angle of 19.5 mrad. The spectrum, Figure 3a, looks very similar to Figure 2a, and shows the bulk LO, volume HPhP, and ZO dipole signals at 194, 175, and 100 meV, respectively. However, the lower energy-loss tail for the 175 meV shoulder, which is associated with BZB phonon modes, is stronger in intensity relative to that in Figure 2a. The dipole and non-dipole components are shown in Figure 3b. The dipole spectrum is remarkably similar to that in Figure 2d which implies that dipole modes are unaffected by an increase in spectrometer collection. The non-dipole component shows a 3-fold increase relative to Figure 2d due to the increased collection angle but is still 10 times weaker than the dipole signal.

*Increasing the Probe Convergence Angle*

An alternative approach to increase Bragg scattering is to make the convergence angle greater than the Bragg angle. The background subtracted spectrum obtained with $\alpha = 33$ mrad and $\beta = 10$ mrad is shown in Figure 4a. The spectral shape is starkly different from Figures 2a and 3a, and there are many additional peaks. The dipole component of the spectrum (Figure 4b) shows good similarity to the dipole contributions recorded with the smaller probe convergence. The main difference being the energy and intensity of the polariton contribution at around 175 meV, and an increase in the ZO peak at 100 meV because of the larger out-of-plane component of momentum transfer enabled by the factor of 3 increase in $\alpha$. The non-dipole component shows three strong peaks at ~157, 126, and 69 meV. The physical origin of the well-defined peaks can be identified by inspection of Figure 2b which shows that they correspond to phonon modes associated with the flat parts of the dispersion curves at or near the BZB. Specifically, the peak between 145 and 160 meV is associated with the TO/LO BZB phonon, the one between 125 and 145 meV is the longitudinal acoustic (LA) BZB phonon while the one between 65 and 75 meV is the transverse acoustic (TA) BZB phonon. All vibrational excitations that are observed in this work are labelled and described with their spectral positions in Table I. Based on this analysis, the non-dipole part of the spectrum is associated with short wavelength modes at the BZB and is a strong signal accounting for 38% of the spectral intensity.

| Label | Mode description | Energy, meV |
|---|---|---|
| LO (long) | LO phonon from BZ center | 193-194 |
| volume HPhP | Volume hyperbolic phonon polariton | 175 |
| edge HPhP | Edge-surface hyperbolic phonon polariton | 184 |
| convolved HPhP | Convolution of volume and edge-surface HPhP | 179 |
| TO/LO (short) | TO/LO BZB phonon | 155-160 |
| LA (short) | LA BZB phonon | 125-134 |
| ZO (long) | ZO phonon from BZ center | 98-100 |
| TA/ZO (short) | TA/ZO BZB phonon | 69 |

Table I. Different vibrational modes excited by the electron beam in [0001] h-BN.

The non-dipole contribution with α = 33 mrad can be further enhanced by increasing β to 40 mrad as shown in Figure 5a. The increase in β not only strengthens the non-dipole contribution but also gives a 3-fold increase in the overall signal intensity relative to Figure 4a. The volume HPhP and bulk dipole peaks at 175 and 194 meV look very similar in shape to that in earlier spectra. The ZO phonon at 100 meV is more than a factor of 4 stronger compared to Figure 4a. The non-dipole spectral component is further increased amounting to 42% of the spectral intensity. The peak energies are approximately similar to Figure 4a, so the same phonon modes are excited. Interestingly, the significant enhancement on the peaks at 69 and 100 meV implies that each is strongly associated with the ZO phonon at the M and K points (69 meV) and at the center of the Brillouin zone (100 meV).

3.3. Localization of Long and Short Wavelength Phonon Modes

The ability to probe the center and boundaries of the Brillouin zone arises because two different scattering mechanisms contribute to the EELS spectrum. The dipole interaction, associated with small scattering angles, only excites the long wavelength modes and is analogous to IR spectroscopy. On the other hand, impact scattering, associated with larger scattering angles, is analogous to neutron scattering and excites all modes in the system including the short wavelength modes. The difference in the average scattering angles associated with dipole and impact scattering should give rise to differences in the delocalization of the signals. Dipole interaction should be more delocalized than impact scattering. To test this hypothesis, the spatial variation profiles of impact and dipole signals across the h-BN flake edge was determined by performing linescans with $\alpha = 33$ mrad and $\beta = 40$ mrad. An ADF image in Figure 6a shows the linescan position relative to the flake; the step size for EELS acquisition is 0.2 nm. The background subtracted vibrational spectra acquired at the depicted positions are shown in Figure 6b (the relative intensity of the convolved HPhP peak (~179 meV) in the transmission spectrum looks different from the volume HPhP peak in Figure 5a due to the changes in the position of the probe with respect to the sample edge). The aloof beam spectrum recorded at the end of the linescan in the vacuum (impact parameter 21 nm) contains no bulk phonon peaks showing only one sharp volume HPhP signal at 175 meV.

Since the dipole and impact peaks are associated with different spectral energy ranges, the localization was explored by plotting integrated intensities contained in different energy windows along the linescan. The background subtracted signal must be normalized by the total signal going into the spectrometer to account for elastic scattering and sample thickness changes along the linescan. Since the ZLP was saturated, the signal was normalized relative to the negative tail of the ZLP, integrated between -250 and -200 meV (no significant energy-gain signal is observed in these spectra). The spatial variation profiles of dipole and impact signals are shown in Figures 7a and 7b, with the simultaneously acquired HAADF signal profile overlaid on both plots for comparison.

The spatial variation profile of the LO (long) signal follows the HAADF profile up to ~10 nm from the edge, which confirms that it is a bulk signal. As the probe approaches the edge of the flake, it falls over a larger distance than the HAADF signal which shows that it is more delocalized. This delocalization is attributed to the begrenzungs effect which is characteristic of a bulk dipole energy-loss signal as the probe moves towards a boundary [23,24]. The signal in vacuum comes from the tail of the aloof signal associated with the volume HPhP at 176 meV (see Figure 6b).

The convolved HPhP signal profile does not follow the HAADF signal profile; it is approximately constant with decreasing flake thickness up to ~10 nm from the edge. This implies that it is more of a surface signal [25]. It then falls rapidly till the probe is ~3 nm from the edge, and then drops more slowly to ~30% of its maximum value at the exact edge position. This shows that the convolved HPhP signal is much more delocalized than the LO (long) and is characteristic of a polariton signal [26]. The inflection in the convolved HPhP profile when the probe is ~3 nm from the edge can be attributed to an increased excitation of the edge-surface HPhP accompanied by a decreased volume HPhP excitation. The spatial variation in vacuum is associated with the aloof volume HPhP signal.

The signals corresponding to the TO/LO (short) and LA (short) phonons mostly follow the ADF intensity as the probe crosses the flake edge indicating that these signals are highly localized (they drop slightly faster than the ADF signal due to a 1 nm surface amorphous layer on the BN flake as seen in Figure S2). The high degree of localization confirms that the TO/LO (short) and LA (short) signals are associated with impact scattering and larger scattering angles. This is consistent with these peaks being associated with excitation of phonons at the BZB.

3.4 Origin of the short wavelength mode enhancement with increased probe convergence

The measurements presented here show that it is possible to acquire energy-loss spectra that probe phonon modes at both the center and boundaries of the Brillouin zone. It has been well established that EELS recorded with on-axis geometry can easily detect long wavelength phonons at the Γ point. It was surprising to find that simply increasing the convergence angle of the electron probe would give a large increase in the contribution to the spectrum from the BZB. The delocalization measurements show the long wavelength modes are probed by the dipole interaction whereas the short wavelength modes are probed by impact scattering. The quantitative difference in the intensity of the BZB peaks are shown in Table II

(normalized to the LO (long) phonon peak which is common to all spectra). The TO/LO (short) and LA (short) in-plane modes increase by factors of 5 – 10 when the probe convergence angle is increased from 10 to 33 mrad. This is much larger than the factor of 2 increase that is associated with increasing the collection angle alone. Another interesting observation was the increase in both the long and short wavelength ZO phonon with increase in both α and β. The ZO phonon involves out-of-plane atomic motion. For a parallel electron beam with α close to zero, the momentum transfer is mostly parallel to the (0001) plane which will not excite the ZO phonon. Increasing the convergence angle provides a large enough component of momentum transfer to significantly increase the excitation probability of this phonon.

| α, β | LO (long) | volume HPhP | TO/LO (short) | LA (short) | ZO (long) | TA/ZO (short) |
|---|---|---|---|---|---|---|
| 10, 10 | 1 | 0.6483 | 0.0762 | 0.0238 | 0.0350 | 0 |
| 10, 40 | 1 | 0.6198 | 0.1227 | 0.0547 | 0.0205 | 0 |
| 33, 10 | 1 | 0.5389 | 0.7422 | 0.1300 | 0.0596 | 0.0453 |
| 33, 40 | 1 | 0.8432 | 0.7093 | 0.2281 | 0.3860 | 0.3446 |

Table II. Integrated intensities of different spectral features as a fraction of the LO (long) signal.

The short wavelength mode enhancement associated with increasing the collection angle is expected and sampling the off-axis scattering is the basis for mapping the phonon dispersion surfaces with EELS [14,15]. It is interesting to consider the origin of the 5 – 10 fold increase in the non-dipole contribution associated with increasing the convergence angle. In electron diffraction from a thin crystal, the strength of a Bragg excitation depends on the excitation error, $s$ [27]. For a typical TEM sample of thickness, $t$, the reciprocal lattice points are replaced with relrods along the beam direction with lengths proportional to $1/t$. For a crystal in a zone axis orientation with α less than the Bragg angle, the excitation error $s \neq 0$ for all rays in the incident probe and consequently the exact Bragg condition is not satisfied. When α exceeds the Bragg angle, there will be a subset of rays in the probe that satisfy the exact Bragg condition. Thus, the Ewald sphere associated with rays in the probe that are incident at the Bragg angle will exactly cut a reciprocal lattice point and give rise to strong Bragg excitation. This is analogous to the 2-beam condition in classical diffraction contrast theory. The enhancement in the intensity between a parallel zone axis illumination and a 2-beam condition is typically a factor of 5 – 10. The strong diffracting condition associated with the zone axis orientation and the exact Bragg condition will also make dynamical effects very important in all but the thinnest of crystals.

These strong diffracting conditions seem to correlate with enhanced spectral contributions from the BZB. To develop a fully quantitative understanding of this effect, it will be necessary to perform extensive calculations from a fully-dynamical phonon scattering theory for electrons. Such calculations are not readily available at present; however, some insights can be gained from the fundamental properties of phonons and the recent vibrational EELS report from Si [28]. It was shown that short wavelength phonons in Si can be detected when large probe convergence angles are employed. Si is a non-polar elemental semiconductor and there is no dipole contribution to the spectrum (whereas h-BN has a strong dipole response). Phonons can be excited either by normal processes, where the scattering wavevector lies within the first BZ, or via Umklapp processes, where the scattering wavevector lies outside the BZ [1]. Recent fully-dynamical phonon scattering calculations for Si suggest that Umklapp processes from the second BZ may be critical in explaining many of the features observed in the on-axis spectrum (P. Rez, private communication). These Umklapp processes are favored under strong diffraction conditions and since the density of states often peaks at the BZB, these are the modes that appear in the spectrum. Further experimental and theoretical work will be required to develop a fundamental quantitative understanding of all the factors which influence the relative strength of the dipole and non-dipole contributions. However, it is clear from the experimental results that when large convergence angles are employed, the forward scattering spectrum contains long wavelength modes from dipole scattering and short wavelength modes from impact scattering.

The difference in the spatial and angular dependence of the dipole and impact scattering processes offers two approaches to separate the short and long wavelength components of the spectrum. Recording two sets of spectra with high and low convergence angles will allow the dipole contributions to be easily identified since they will be approximately constant in each spectrum. Alternatively, spatial differencing techniques would also allow the dipole and impact contributions to be separated since the dipole signal will be much more delocalized. From a practical point of view, these two approaches have considerable utility. Spectral peaks can be associated with long or short wavelength phonons without the need for electron scattering calculations or detailed knowledge of the dispersion curves of the system. The latter point is particularly important since the motivation for performing experiments in the STEM is to probe local atomic scale heterogeneities such as defects or interfaces and determine the vibrational response. On a modern STEM, changing the convergence angle or spectral mapping is easily accomplished allowing rapid interpretation of the spectral features.

4. Conclusion

We have explored how dipole and impact electron scattering can be exploited to probe long and short wavelength phonon modes in crystals using electron energy-loss spectroscopy in a scanning transmission electron microscope. By varying the incident electron probe convergence and spectrometer collection angles, the relative contribution of dipole and impact scattering to the recorded spectrum can be varied. Specifically, optimum conditions for exciting phonons dominated by impact scattering and detecting them with a conventional on-axis STEM detection geometry were investigated in h-BN, a polar crystal. Employing a probe convergence semi-angle less than the (1000) Bragg angle of h-BN results in an energy-loss spectrum dominated by dipole scattering giving strong spectral contributions from long wavelength phonons and polaritons from the central region of the Brillouin zone. Increasing the probe convergence to include the Bragg angle enhances the impact scattering contribution yielding a spectrum containing not only long wavelength modes but also a significant contribution from short wavelength modes associated with the BZB. These impact signals are much more localized than the dipole signals. As the probe moves towards the BN flake edge, the dipole profile associated with the long wavelength bulk LO phonon shows a characteristic attenuation over a 10 nm length scale due to the begrenzungs effect. However, the LO phonon signal is more localized than the HPhP signal which extends many tens of nanometers into the vacuum. The impact signal profiles are sharper and trace the HAADF profiles, thereby confirming their high spatial resolution.

Probing long and short wavelength phonons simultaneously with STEM EELS offers vibrational information on materials similar to that available from combining IR absorption and neutron scattering spectroscopies. A huge advantage of the on-axis vibrational STEM EELS approach is that we can probe short wavelength modes rapidly and with high spatial resolution. The condition required for high spatial resolution in STEM, namely a convergence angle which exceeds the relevant Bragg angle, is also found to be the condition that enhances the short wavelength phonon contribution to the energy-loss spectrum. This ability to probe short wavelength modes rapidly with high spatial resolution will help establish a direct connection between vibrational excitations and local atomic level defects and structural heterogeneities such as surfaces and interfaces in materials.

5. Acknowledgements

The authors acknowledge the insightful comments from Prof. Peter Rez on the manuscript. They also acknowledge the financial support from the US National Science Foundation (grant no. CHE-1508667) and the use of (S)TEM at John M. Cowley Center for High Resolution Electron Microscopy in the Eyring Materials Center at Arizona State University.

6. References

[1] N. W. Ashcroft and N. D. Mermin, *Solid State Physics* (Cengage Learning, 2011).
[2] A. A. Govyadinov, A. Konečná, A. Chuvilin, S. Vélez, I. Dolado, A. Y. Nikitin, S. Lopatin, F. Casanova, L. E. Hueso, J. Aizpurua, and R. Hillenbrand, *Probing Low-Energy Hyperbolic Polaritons in van Der Waals Crystals with an Electron Microscope*, Nat. Commun. **8**, 95 (2017).
[3] H. Ibach and D. L. Mills, *Electron Energy Loss Spectroscopy and Surface Vibrations* (Academic Press, New York, 1982).
[4] B. S. Hudson, D. G. Allis, S. F. Parker, A. J. Ramirez-Cuesta, H. Herman, and H. Prinzbach, *Infrared, Raman, and Inelastic Neutron Scattering Spectra of Dodecahedrane: An I h Molecule in T h Site Symmetry*, J. Phys. Chem. A **109**, 3418 (2005).
[5] P. C. H. Mitchell, *Vibrational Spectroscopy with Neutrons: With Applications in Chemistry, Biology, Materials Science and Catalysis*, Vol. 3 (World Scientific, 2005).
[6] J. Serrano, A. Bosak, R. Arenal, M. Krisch, K. Watanabe, T. Taniguchi, H. Kanda, A. Rubio, and L. Wirtz, *Vibrational Properties of Hexagonal Boron Nitride: Inelastic X-Ray Scattering and* Ab Initio *Calculations*, Phys. Rev. Lett. **98**, 095503 (2007).
[7] R. Haworth, G. Mountjoy, M. Corno, P. Ugliengo, and R. J. Newport, *Probing Vibrational Modes in Silica Glass Using Inelastic Neutron Scattering with Mass Contrast*, Phys. Rev. B **81**, 060301 (2010).
[8] G. Bracco and B. Holst, *Surface Science Techniques* (Springer Science & Business Media, 2013).
[9] A. Hartschuh, N. Anderson, and L. Novotny, *Near-field Raman Spectroscopy Using a Sharp Metal Tip*, J. Microsc. **210**, 234 (2003).
[10] F. Huth, A. Govyadinov, S. Amarie, W. Nuansing, F. Keilmann, and R. Hillenbrand, *Nano-FTIR Absorption Spectroscopy of Molecular Fingerprints at 20 Nm Spatial Resolution*, Nano Lett. **12**, 3973 (2012).
[11] S. J. Pennycook and P. D. Nellist, editors, *Scanning Transmission Electron Microscopy* (Springer New York, New York, NY, 2011).
[12] O. L. Krivanek, T. C. Lovejoy, N. Dellby, T. Aoki, R. W. Carpenter, P. Rez, E. Soignard, J. Zhu, P. E. Batson, M. J. Lagos, R. F. Egerton, and P. A. Crozier, *Vibrational Spectroscopy in the Electron Microscope*, Nature **514**, 209 (2014).
[13] T. Miyata, M. Fukuyama, A. Hibara, E. Okunishi, M. Mukai, and T. Mizoguchi, *Measurement of Vibrational Spectrum of Liquid Using Monochromated Scanning Transmission Electron Microscopy - Electron Energy Loss Spectroscopy*, Microscopy **63**, 377 (2014).
[14] F. S. Hage, R. J. Nicholls, J. R. Yates, D. G. McCulloch, T. C. Lovejoy, N. Dellby, O. L. Krivanek, K. Refson, and Q. M. Ramasse, *Nanoscale Momentum-Resolved Vibrational Spectroscopy*, Sci. Adv. **4**, eaar7495 (2018).
[15] R. Senga, K. Suenaga, P. Barone, S. Morishita, F. Mauri, and T. Pichler, *Position and Momentum Mapping of Vibrations in Graphene Nanostructures*, Nature **573**, 247 (2019).
[16] O. L. Krivanek, P. D. Nellist, N. Dellby, M. F. Murfitt, and Z. Szilagyi, *Towards Sub-0.5Å Electron Beams*, Ultramicroscopy **96**, 229 (2003).
[17] O. L. Krivanek, J. P. Ursin, N. J. Bacon, G. J. Corbin, N. Dellby, P. Hrncirik, M. F. Murfitt, C. S. Own, and Z. S. Szilagyi, *High-Energy-Resolution Monochromator for Aberration-Corrected Scanning Transmission Electron Microscopy/Electron Energy-Loss Spectroscopy*, Philos. Trans. R. Soc. Math. Phys. Eng. Sci. **367**, 3683 (2009).
[18] B. D. Levin, K. Venkatraman, D. M. Haiber, K. March, and P. A. Crozier, *Background Modelling for Quantitative Analysis in Vibrational EELS*, Microsc. Microanal. **25**, 674 (2019).
[19] K. Iakoubovskii, K. Mitsuishi, Y. Nakayama, and K. Furuya, *Thickness Measurements with Electron Energy Loss Spectroscopy*, Microsc. Res. Tech. **71**, 626 (2008).
[20] R. H. Ritchie, *Plasma Losses by Fast Electrons in Thin Films*, Phys. Rev. **106**, 874 (1957).

[21] R. Geick, C. H. Perry, and G. Rupprecht, *Normal Modes in Hexagonal Boron Nitride*, Phys. Rev. **146**, 543 (1966).

[22] A. Poddubny, I. Iorsh, P. Belov, and Y. Kivshar, *Hyperbolic Metamaterials*, Nat. Photonics **7**, 948 (2013).

[23] R. F. Egerton, *Electron Energy-Loss Spectroscopy in the Electron Microscope* (Springer US, Boston, MA, 2011).

[24] K. Venkatraman, P. Rez, K. March, and P. A. Crozier, *The Influence of Surfaces and Interfaces on High Spatial Resolution Vibrational EELS from SiO2*, Microscopy **67**, i14 (2018).

[25] R. F. Egerton, K. Venkatraman, K. March, and P. A. Crozier, *Properties of Dipole-Mode Vibrational Energy Losses Recorded From a TEM Specimen*, Microsc. Microanal. 1 (2020).

[26] A. Konečná, K. Venkatraman, K. March, P. A. Crozier, R. Hillenbrand, P. Rez, and J. Aizpurua, *Vibrational Electron Energy Loss Spectroscopy in Truncated Dielectric Slabs*, Phys. Rev. B **98**, 205409 (2018).

[27] D. B. Williams and C. B. Carter, *Transmission Electron Microscopy: A Textbook for Materials Science*, 2nd ed (Springer, New York, 2008).

[28] K. Venkatraman, B. D. A. Levin, K. March, P. Rez, and P. A. Crozier, *Vibrational Spectroscopy at Atomic Resolution with Electron Impact Scattering*, Nat. Phys. **15**, 1237 (2019).

Figures and figure captions

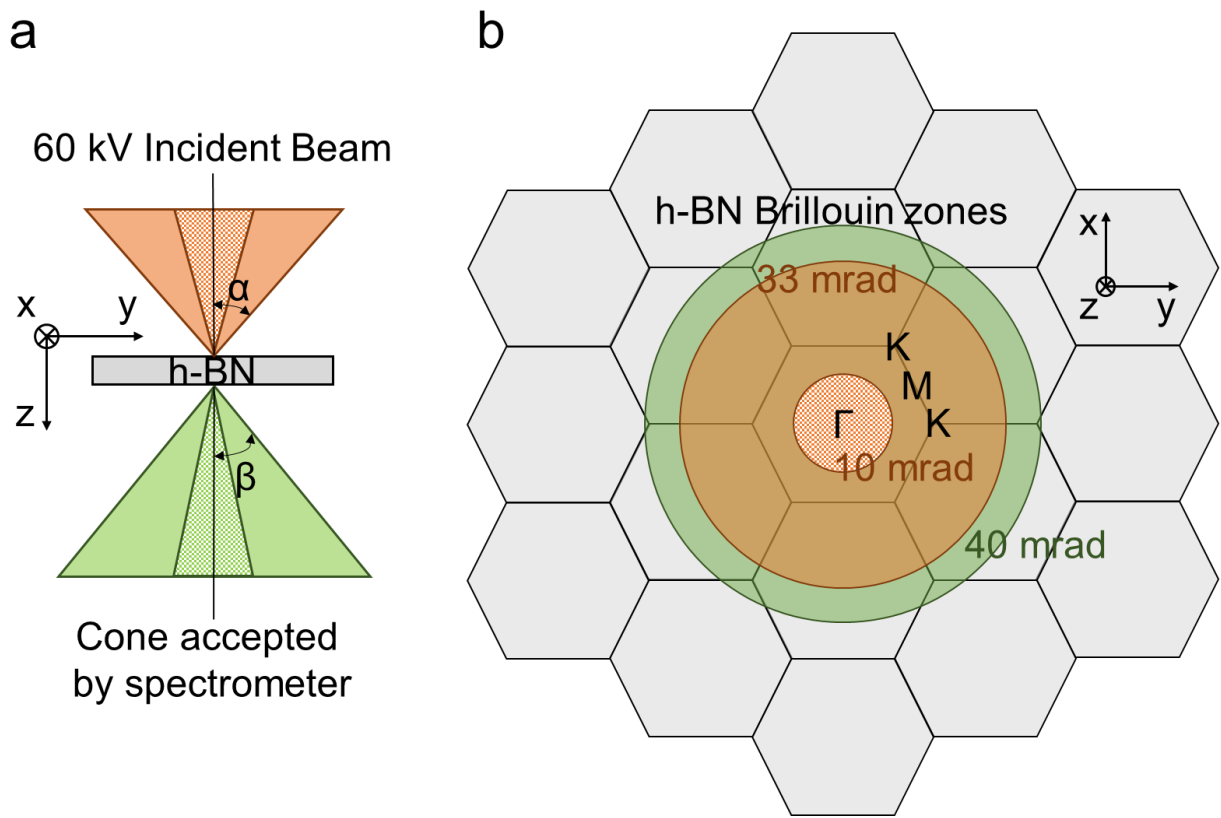

**Figure 1.** (a) STEM EELS geometry employed during data acquisition. The solid orange cone above the sample denotes 33 mrad convergence, while the shaded orange cone denotes 10 mrad convergence. The solid green cone below the sample represents 40 mrad collection, while the shaded green cone represents 10 mrad collection. (b) Circles representing the different convergence and collection semi-angles employed are overlaid on the hexagonal Brillouin zones of [0001] h-BN.

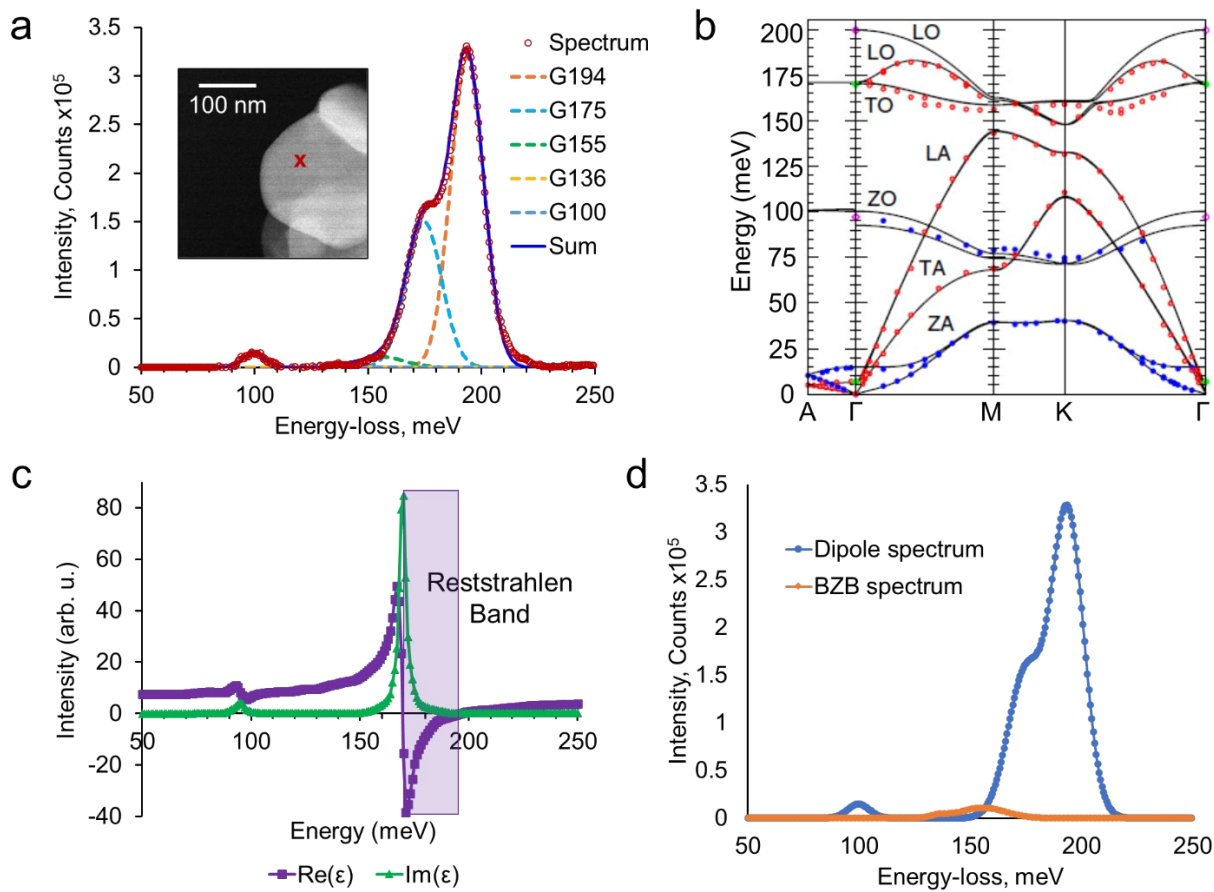

**Figure 2.** (a) Background subtracted vibrational spectrum with 10 mrad convergence and collection semi-angles. Gaussians with peaks associated with flat parts of the phonon dispersion curves from bulk h-BN [6] in (b) were fit to the spectrum. (Inset in a) ADF image showing probe position on the h-BN flake. (c) In-plane dielectric function of h-BN showing the Restrahlen band between 170 and 196 meV [21]. (d) Dipole and BZB contribution to the spectrum in (a).

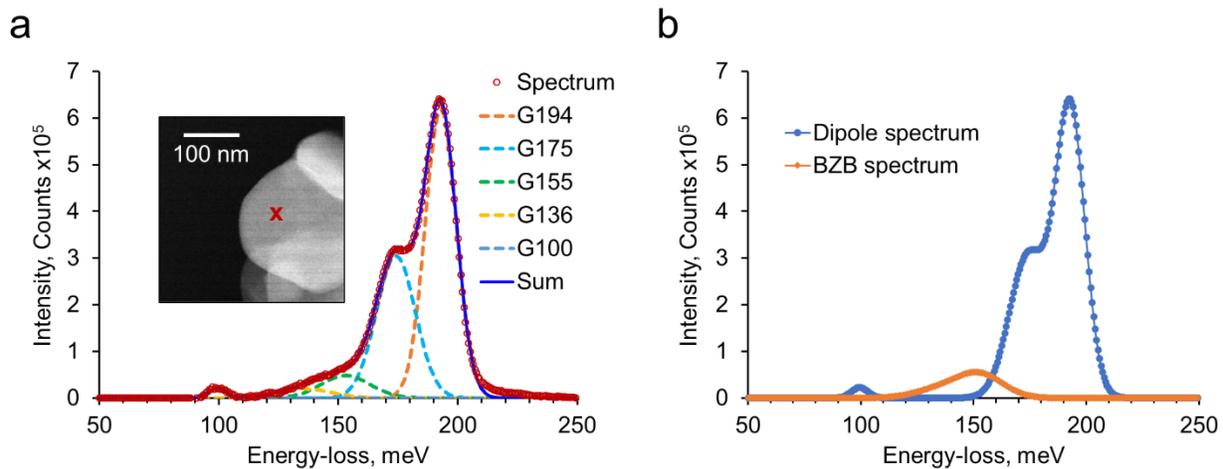

**Figure 3.** (a) Background subtracted vibrational spectrum with 10 mrad convergence and 40 mrad collection semi-angles. Gaussian fitting was also performed as shown. (Inset) ADF image of the flake showing the probe position. (b) Dipole and BZB contribution to the spectrum in (a).

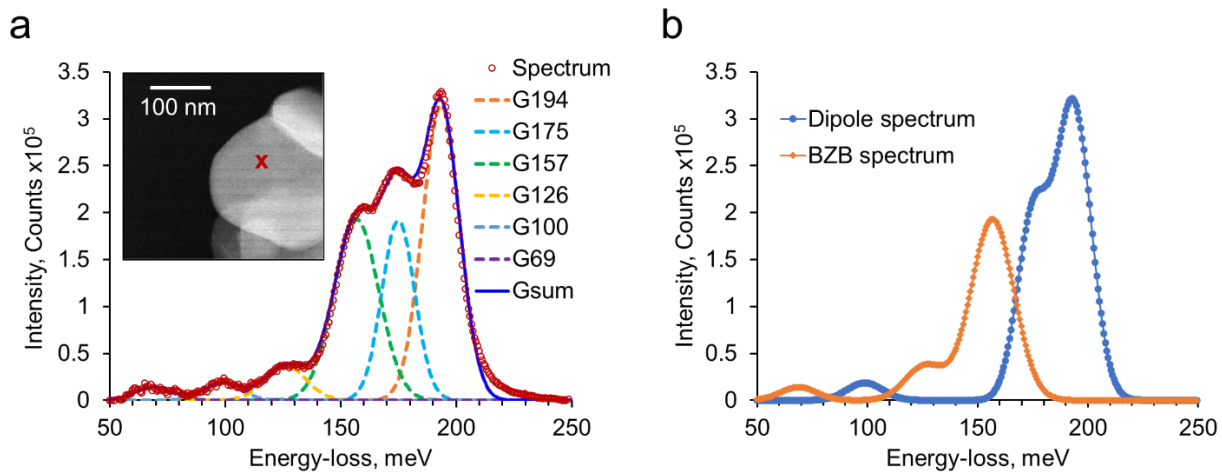

**Figure 4.** (a) Background subtracted vibrational spectrum with 33 mrad convergence and 10 mrad collection semi-angles. Gaussian fitting was also performed as shown. (Inset) ADF image of the flake showing the probe position. (b) Dipole and BZB contribution to the spectrum in (a).

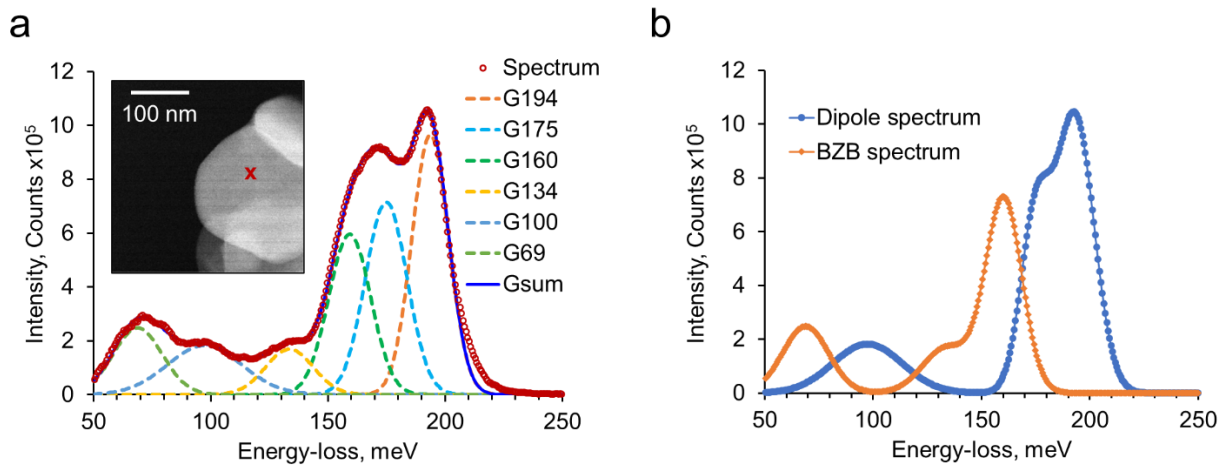

**Figure 5.** (a) Background subtracted vibrational spectrum with 33 mrad convergence and 40 mrad collection semi-angles. Gaussian fitting was also performed as shown. (Inset) ADF image of the flake showing the probe position. (b) Dipole and BZB contribution to the spectrum in (a).

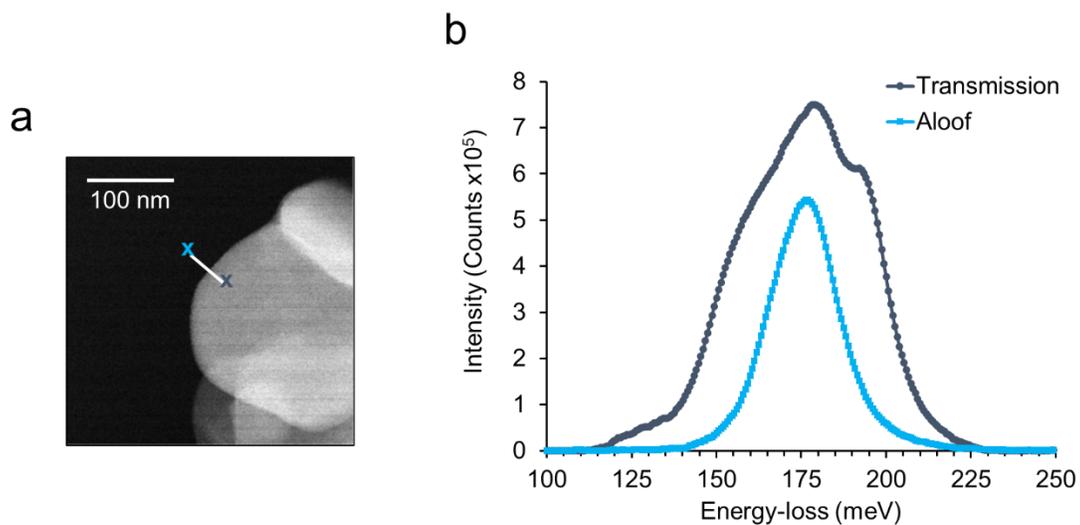

**Figure 6.** (a) ADF image showing the position of the linescan across the h-BN flake edge. Linescan length = 50 nm, 250 pixels. (b) Background subtracted transmission and aloof beam spectra at the extremes of the linescan with 33 mrad convergence and 40 mrad collection semi-angles. Impact parameter (distance of the probe from the flake edge) for the transmission spectrum was 29 nm, while that for the aloof spectrum was 21 nm.

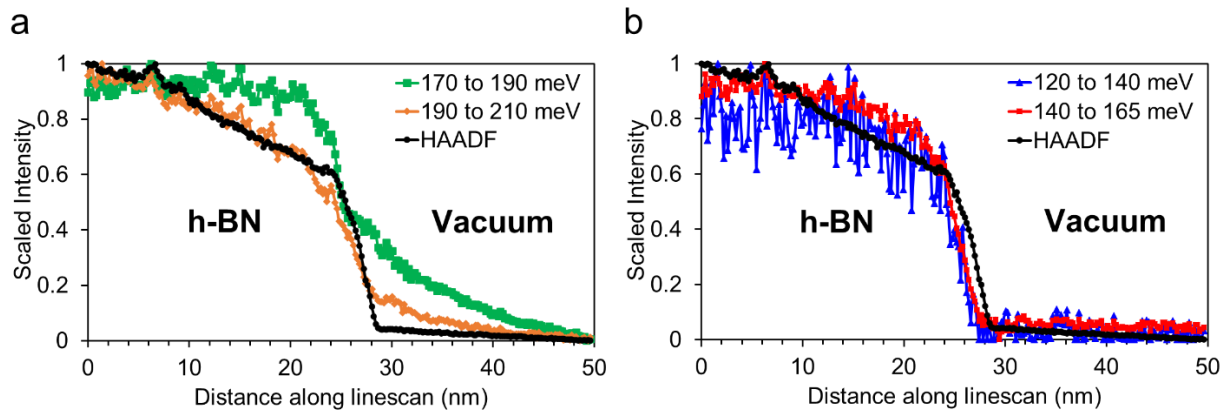

**Figure 7.** (a) Spatial variation of dipole signals along the linescan overlaid on the HAADF profile. (b) Spatial variation of impact signals along the linescan overlaid on the HAADF profile.

Supplementary Figures

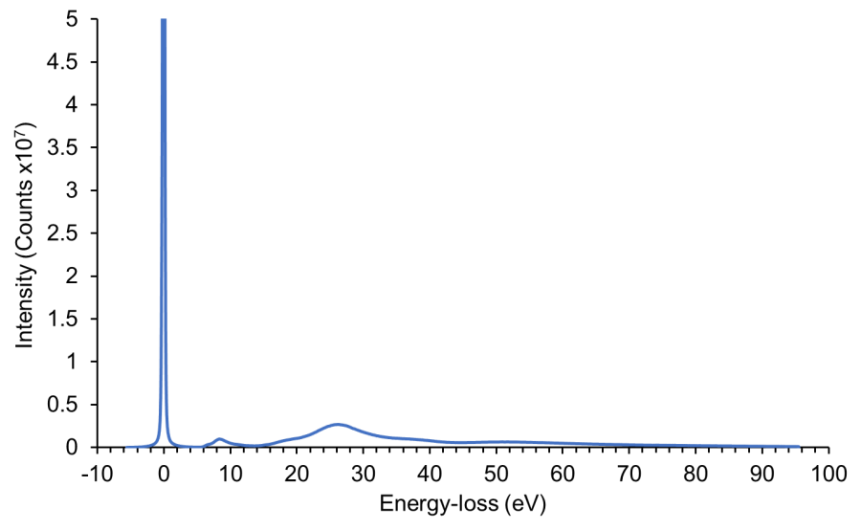

**Figure S1.** Low energy-loss spectrum from h-BN for estimating flake thickness. $t/\lambda = 0.68$ for the h-BN flake shown in earlier ADF images. The flake thickness t is then estimated to be ~50 nm, given the inelastic mean free path for valence electron scattering was calculated to be 75 nm [19].

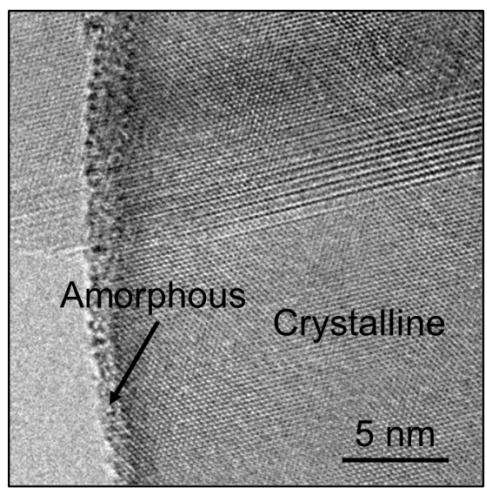

**Figure S2.** Bright-field (BF) TEM image of a h-BN flake showing an approximately 1 nm thick surface amorphous layer.